\documentclass[a4paper]{article}
\usepackage{ASVspoof2021}
\usepackage{epsfig,amssymb,amsmath}
\ninept
\usepackage{booktabs}
\usepackage{balance,subcaption,xcolor}
\usepackage[symbol]{footmisc}
\usepackage{hyperref}
\usepackage{ctable}
\title{Raw Differentiable Architecture Search \\
for Speech Deepfake and Spoofing Detection}

\name{Wanying Ge, Jose Patino, Massimiliano Todisco and Nicholas Evans}

\address{EURECOM, Sophia Antipolis, France \\
{\small \tt lastname@eurecom.fr} }

\begin{document}
\maketitle

\begin{abstract}
    End-to-end approaches to anti-spoofing, especially those which operate directly upon the raw signal, are starting to be competitive with their more traditional counterparts.  Until recently, all such approaches consider only the learning of network parameters; the network architecture is still hand crafted.  This too, however, can also be learned. Described in this paper is our attempt to learn automatically the network architecture of a speech deepfake and spoofing detection solution, while jointly optimising other network components and parameters, such as the first convolutional layer which  operates on raw signal inputs. The resulting raw differentiable architecture search system delivers a tandem detection cost function score of 0.0517 for the ASVspoof 2019 logical access database, a result which is among the best single-system results reported to date.
\end{abstract}

\section{Introduction}
    End-to-end (E2E) solutions are attracting growing attention across a broad range of speech processing tasks~\cite{ravanelli2018sincnet, luo2019Conv-TesNet, peter2021e2ekeyword}. In contrast to the more common approach whereby front-end feature extraction and the back-end classifier or network are separately optimised, E2E solutions allow for pre-processing and post-processing components to be combined within a single network.  With both components being encapsulated within a single model,  front-end and back-end components can be jointly optimised. In this case the front-end might have a better chance of capturing more discriminative information for the task in hand~\cite{dinkel2017cldnns,tak2021rawnet2,hua2021TSSDNet}, whereas the back-end might be able to function more effectively upon the information to produce more reliable scores.
    
    Many solutions to anti-spoofing for automatic speaker verification have focused upon the design of deep neural network (DNN) based back-end classifiers.  Most combine fixed, hand-crafted features, usually in the form of some spectro-temporal decomposition~\cite{todisco2016cqcc,todisco2019asvspoofbaselines}, with a convolutional neural network (CNN) to learn higher-level representations. The literature shows that the use of specially designed network modules~\cite{lavrentyeva2019lcnn, li2020res2net, tak2021GAT} and loss functions~\cite{wang2021comparative, chen2020generalization, zhang2021oneclass} generally leads to better performing models. Still, their potential is fundamentally dependent upon the information captured in the initial features; information lost in initial feature extraction cannot be recovered.  Several works have also shown that the performance of a given model can vary substantially when fed with different features~\cite{lavrentyeva2019lcnn, li2020res2net,wang2021comparative}.  These observations point toward the importance of learning and optimising not just the higher-level representation, but also the initial features, in unison with the classifier.

    E2E solutions have been a focus of our research group for some time~\cite{valenti2018end}.  Fundamental to this pursuit is operation upon the raw signal.  A recent attempt~\cite{tak2021rawnet2} adopted the RawNet2 architecture~\cite{jung2019rawnet, jung2020rawnet2}.  Using a bank of sinc-shaped filters, it operates directly upon the raw audio waveform through time-domain convolution, with the remaining network components being optimised in the usual way.  Results show that systems that use automatically learned features are competitive and complementary to systems that use hand crafted features. While these findings are encouraging, improvements to performance are perhaps only modest. Despite the emphasis upon the E2E learning of both features and classifier, one aspect of our model remains hand-crafted~\cite{tak2021rawnet2}.  This is also the case for every E2E solution proposed thus far~\cite{dinkel2017cldnns, hua2021TSSDNet,jung2019rawnet}; the network \emph{parameters} are learned, but the network \emph{architecture} is still hand-crafted.
    
    We have hence explored automatic approaches to learn the network architecture as well.  Our first attempt~\cite{ge2021pc-darts} was based upon a specific variant of differentiable architecture search~\cite{liu2018darts} known as partially-connected differentiable architecture search (PC-DARTS)~\cite{xu2019pcdarts}.  Architecture search is performed using a pair of core network components referred to as cells.  Cells are defined by both architecture parameters and network parameters, both of which are jointly optimised during the first of two stages referred to as the \emph{architecture search} stage.
    
    We showed~\cite{ge2021pc-darts} that PC-DARTS learns more compact models that are nonetheless competitive with the state of the art.  As the very first attempt to harness the power of differentiable architecture search for anti-spoofing, this work was performed with hand-crafted features.  Our latest work has hence sought to combine architecture search with fully E2E learning.  In this paper, we present Raw PC-DARTS. It is the first E2E speech deepfake and spoofing detection solution which operates directly upon the raw waveform while allowing for the joint optimisation of both the network architecture and network parameters.
    
    The remainder of the paper is organised as follows. Section~2 introduces the related works. The proposed system is described in Section~3. Reported in Sections~4 and~5 are our experiments and results. Our conclusions are reported in Section~6.
\vspace{-0.37cm}
\section{Related works}
    
    In this section we introduce the two stages of DARTS-based NAS solutions~\cite{liu2018darts,xu2019pcdarts, chen2019pdarts}, namely the architecture search stage using partial connections~\cite{xu2019pcdarts} and the train from scratch stage.
    
    The architecture search stage aims to determine a base component or building block upon which the full model is constructed.  This base component is referred to as a cell.  The term \emph{architecture} refers to the configuration of nodes and interconnections within the cell. 
    
    \begin{figure}[!t]
         \centering
         \includegraphics[width=\columnwidth]{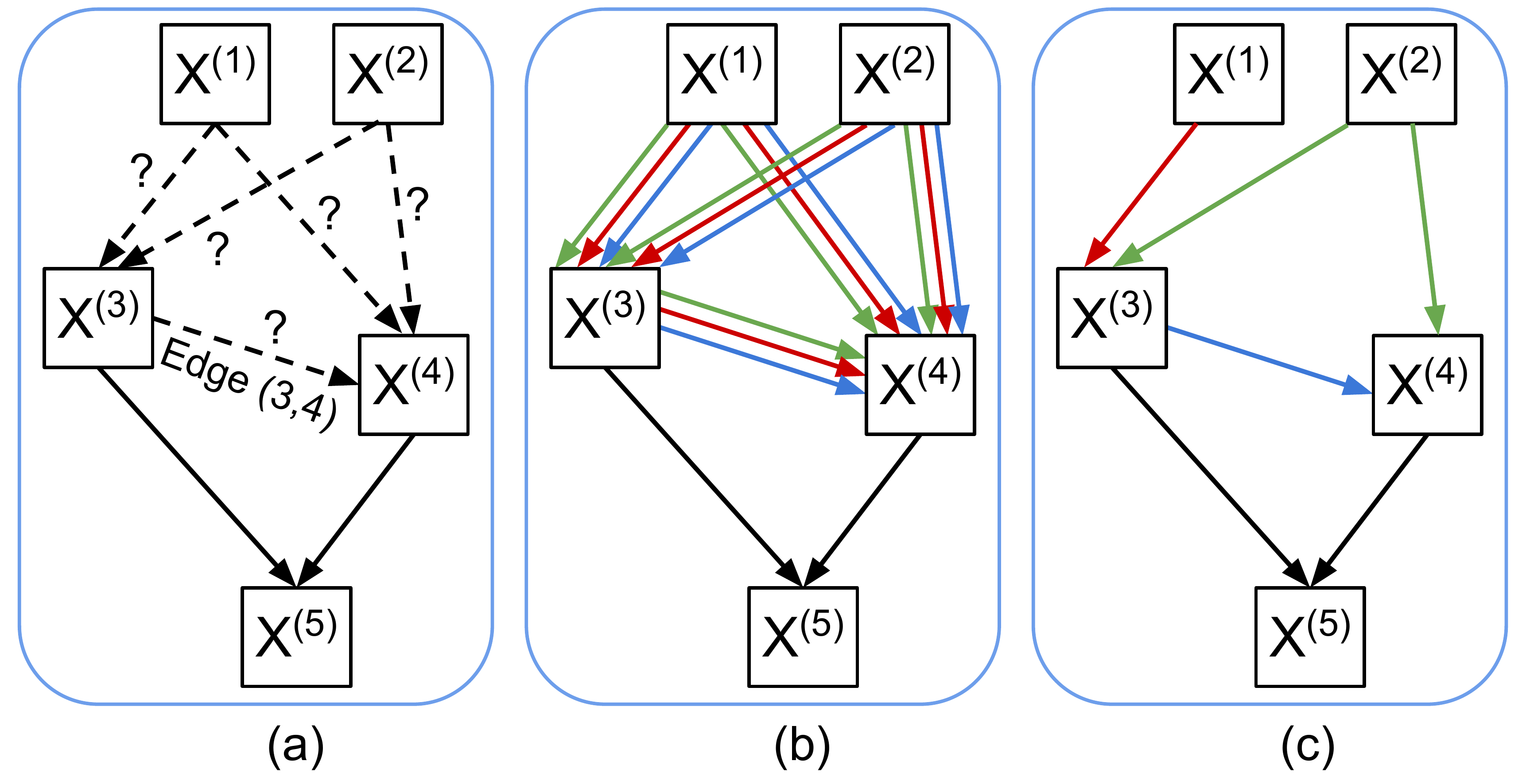}
         \caption{An illustration of architecture search: (a)~a neural cell with \(N=5\) nodes; (b) an illustration of the candidate operations performed on each edge that are optimised during architecture search; (c)~resulting optimised cell with \(2\) inputs to each intermediate node.}
         \label{fig:search stage}
         \end{figure}
    
    As shown in Fig.~\ref{fig:search stage}, each cell has a pair of inputs:  \(\mathbf{x}^{\left(1\right)}\) and \(\mathbf{x}^{\left(2\right)}\).  Cells have a single output, denoted by \(\mathbf{x}^{\left(N\right)}\) ($N=5$ in Fig.~\ref{fig:search stage}). Nodes in between the inputs and output are referred to as intermediate nodes ($\mathbf{x}^{\left(3\right)}$ and $\mathbf{x}^{\left(4\right)}$
    in Fig.~\ref{fig:search stage}). Architecture search involves the selection of candidate operations $o$ from search space $\mathcal{O}$ (solid coloured lines).  Operations between intermediate nodes and the output are fixed to concatenation operations (solid black lines). Each intermediate node is calculated according to:
    
    \begin{equation} \label{eq:node}
        \mathbf{x}^{\left(j\right)}=\sum_{i<j}o^{\left(i,j\right)}\left(\mathbf{x}^{\left(i\right)}\right)
        \end{equation}
    where $o^{\left(i,j\right)}$ is the operation performed on edge $(i,j)$ connecting $\mathbf{x}^{\left(i\right)}$ to $\mathbf{x}^{\left(j\right)}$. 
    During the architecture search stage, the full set of operation candidates are active, with each being assigned a weight
    $\alpha_o^{\left(i,j\right)}$.
    The operation performed on edge $(i,j)$ is then defined as:
    \begin{equation} \label{eq:mixed_op}
        \Bar{o}^{\left(i,j\right)}\left(\mathbf{x}^{\left(i\right)}\right) = \sum_{o \in \mathcal{O}} \frac{\exp \left(\alpha_o^{\left(i,j\right)}\right)}{\sum_{o' \in \mathcal{O}} \exp \left(\alpha_{o'}^{\left(i,j\right)}\right)} \, o\left(\mathbf{x}^{\left(i\right)}\right)
        \end{equation}
    When architecture search is complete, only the single operation with the highest weight $\alpha_o^{\left(i,j\right)}$ is retained. All other operations are discarded; their weights are set to zero.
    
    Because the set of operation weights $\boldsymbol\alpha=\{\alpha^{\left(i,j\right)}\}$ are learnable, the search process is a bi-level optimisation problem.  We seek to determine the weight parameters $\boldsymbol\alpha$ which minimise the validation loss $L_{val}$, while the set of network parameters $\boldsymbol\omega$ is determined by minimising the training loss \(L_{train}(\boldsymbol\omega,\boldsymbol\alpha)\):
        \begin{equation}
        \begin{aligned}
        &\min_{\boldsymbol\alpha} L_{val}(\boldsymbol{\omega}^*,\boldsymbol\alpha) \\
        &\text{s.t.}\;\;\boldsymbol{\omega}^* = \underset{\boldsymbol\omega}{\operatorname{argmin}}\; L_{train}(\boldsymbol\omega,\boldsymbol\alpha)
        \end{aligned}
        \end{equation}
    
    The bi-level optimisation process is demanding in terms of GPU memory and computation. Partial channel connections~\cite{xu2019pcdarts} were proposed as a solution to improve efficiency, reducing demands on both computation and memory.  A binary masking operator $\mathbf{S}^{\left(i,j\right)}$ is used in partially connected (PC) DARTS in order to reduce the complexity of (\ref{eq:mixed_op}). The number of active channels in $\mathbf{x}^{\left(i\right)}$ is reduced through either selection (marked as $\mathbf{S}^{\left(i,j\right)}=1$) or masking (marked as $\mathbf{S}^{\left(i,j\right)}=0$) according to:
    \begin{multline}
    \label{eq:partial connection}
    \Bar{o}^{\left(i,j\right)}\left(\mathbf{x}^{\left(i\right)}\right) =
    \sum_{o \in \mathcal{O}} \frac{\exp{\left(\alpha_o^{\left(i,j\right)}\right)}}{\sum_{o' \in \mathcal{O}} \exp{\left(\alpha_{o'}^{\left(i,j\right)}\right)}}\, o\left(\mathbf{S}^{\left(i,j\right)} \odot \mathbf{x}^{\left(i\right)}\right)\\ + \left(1 - \mathbf{S}^{\left(i,j\right)}\right) \odot \mathbf{x}^{\left(i\right)}
    \end{multline}
    where $\odot$ indicates element wise multiplication. In practice, only a number $1/K_C$ of channels in $\mathbf{x}^{\left(i\right)}$ are selected. The factor $K_C$ is set as a hyper-parameter and acts to trade off  performance (smaller $K_C$) for efficiency (larger~$K_C$).
    
\begin{figure}[!t]
         \centering
         \includegraphics[width=\columnwidth]{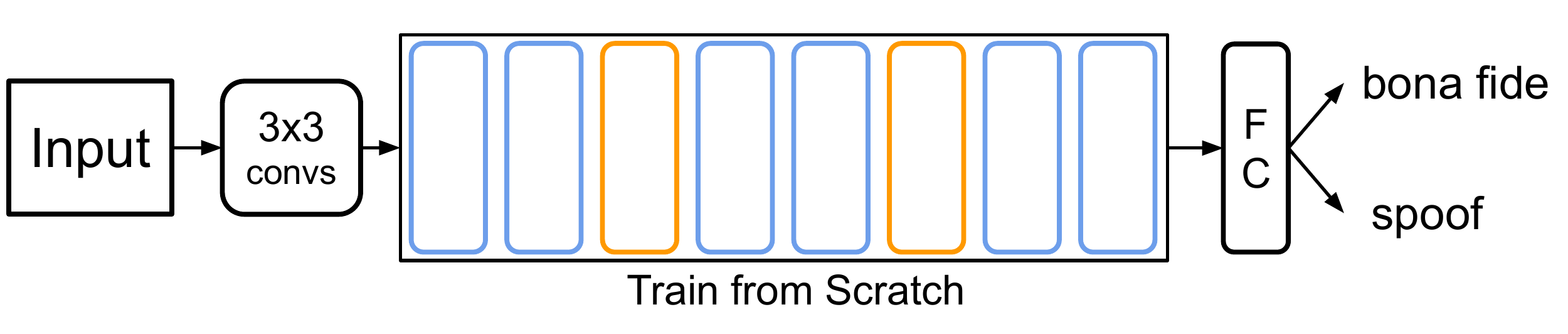}
         \caption{An illustration of train from scratch stage: normal cells (blue) and reduction cells (yellow) are stacked to form a deeper network.}
         \label{fig:train from scratch stage}
         \end{figure}

    After architecture search, the cells 
    are concatenated multiple times (Fig.~\ref{fig:train from scratch stage}) in similar fashion to a ResNet architecture to produce a deeper, more complex model before being further optimised.
    
        \vspace{0.1cm}
\section{Raw PC-DARTS}
        \vspace{0.1cm}

    In this section, we describe the proposed Raw PC-DARTS approach.  The model structure is detailed in Table~\ref{tab:model_structure}. We describe the bank of front-end sinc filters, the application of filter masking, the modifications made to the back-end classifier design and base cell architecture, embedding extraction and the loss function.
    
    \begin{table}[!t]
        \small
	    \centering
    	\caption{The proposed network structure. Each cell receives outputs of its two previous cells/layers. Conv(\(k\), \(s\), \(c\)) stands for a convolutional operation with kernel size \(k\), stride \(s\) and output channel \(c\). BN refers to batch normalisation.}
        \setlength\tabcolsep{3.5pt}
	    \begin{tabular}{ *{3}{c}}
	    
	        \hline
	    Layer & Input:{64000} samples & Output shape\\
	        \hline
	     & Conv(128, 1, 64)  & \\
	   Sinc Filters & Maxpooling(3) & (21290, 64)\\
	    & BN \& LeakyReLU\\
	        \hline
	     & Conv(3, 2, 64) & \\
	    Conv\_1 & BN \& LeakyReLU&(10645, 64)\\
	        \hline
	    Normal Cells & 
	    {$ \left \{ 
	    \begin{array}{c}
	    \text{BN \& LeakyReLU} \\
	    \text{Operations}\\
	    \text{Maxpooling(2) } \\
	    \end{array} \right \} 
	    \times 2$} & (2661,256) \\
	        \hline
	    & BN \& LeakyReLU & \\
	    Expand Cell & Operations & (1330, 512)\\
	    & Maxpooling(2) &\\
	        \hline
	    Normal Cells & 
    	{$ \left \{ 
	    \begin{array}{c}
	    \text{BN \& LeakyReLU} \\
	    \text{Operations}\\
	    \text{Maxpooling(2) } \\
	    \end{array} \right \} 
	    \times 2$} & (332, 512) \\
	        \hline
        & BN \& LeakyReLU & \\
	    Expand Cell & Operations & (166, 1024)\\
	    & Maxpooling(2) &\\
	        \hline
	    Normal Cells & 
	    {$ \left \{ 
	    \begin{array}{c}
	    \text{BN \& LeakyReLU} \\
	    \text{Operations}\\
	    \text{Maxpooling(2) } \\
	    \end{array} \right \} 
	    \times 2$} & (41, 1024) \\
	        \hline
	    GRU & GRU(1024) & (1024)\\
	        \hline
	    Embedding & FC(1024) & (1024)\\
	        \hline
	    Output Score & P2SActivationLayer(2) & (2)\\
	        \hline
	        \label{tab:model_structure}
	\end{tabular}
    \end{table}

    \subsection{Sinc filters and masking}
    \label{subsec:sinc_filter}
    
    The input waveform is fixed to a duration of 4 seconds (\(16000 \times 4\) samples) either by concatenation or truncation of  source audio data. Feature extraction is performed using a set of $C$ sinc filters~\cite{ravanelli2018sincnet}. Each filter performs time-domain convolution upon the input waveform. The impulse response of each filter is defined according to:
    \begin{equation}
        g[n,f_{1}, f_{2}]=2f_{2}sinc(2\pi f_{2}n)-2f_{1}sinc(2\pi f_{1}n)
        \label{equa:sinc}
    \end{equation}
    where \(f_{1}\) and \(f_{2}\) are the cut in and cut off frequencies, and \(sinc(x)=sin(x)/x\) is the sinc function. 
    The cut in and cut off frequencies can be initialised according to any given frequency scale.  Both $f_1$ and $f_2$ are learnable model parameters, though we consider both learnable and fixed configurations. 
    \vspace{0.05cm}

    Filter masking is applied to mask a number of the sinc filters. This is akin to channel drop-out~\cite{cai2019drop-out1, hou2019drop-out2} and frequency masking~\cite{chen2020generalization,park2019specaugment, wang2021specaugmentpp} and acts to encourage the learning of better generalised representations. In practice, sinc filters in the range of \([C_{1}, C_{2})\) are set to zero (masked), where \(C_{1}\) is the first masked filter selected at random and \(C_{2}=C_{1}+f\). The number of masked filters \(f\) is chosen from a uniform distribution \([0, F)\), where \(F\) is a pre-defined maximum value. After $f$ is generated, $C_{1}$ is then chosen from a uniform distribution \([0, C-f)\).

    \subsection{Search space and cell architectures}
    
    In contrast to the approach described in~\cite{ge2021pc-darts} where input features can be seen as a 2D image, operations in Raw PC-DARTS are performed directly upon the raw time-domain waveform.  Thus, the search space $\mathcal{O}$ is designed based on 1D convolutional operations, which includes: standard convolution and dilated convolution with kernel size \{3, 5\}; max pooling and average pooling with kernel size \{3\}; skip connections; no connections.
    
    The original DARTS approach searches for the architectures of two types of cells, namely a normal cell and a reduction cell. The model is formed by stacking these cells sequentially, with the reduction cells being placed at $\frac{1}{3}$ and $\frac{2}{3}$ of the total network depth. While the normal cell preserves the feature map dimension, the reduction cell reduces the dimension by one-half, while the number of channels is doubled.  A global average pooling layer is then used after the stacked network to extract embeddings.
    
    This stacked cell design works well for spectro-temporal representations since their dimensions are close to those used typically in image classification tasks to which DARTS was first applied~\cite{deng2009ImageNet, krizhevsky2009CIFAR-10}. For speech classification tasks and for solutions that operate upon raw inputs, however, the feature dimension remains large at the stacked cell output and the use of global pooling will result in the substantial loss of information.  While a larger number of reduction cells can be added manually to help reduce the feature dimension, this would defeat the purpose of searching the architecture automatically. The introduction of each additional reduction cell also doubles the number of channels, which in turn increases prohibitively both computational complexity as well as demands upon GPU memory.
    
    To address this problem in Raw PC-DARTS, we apply maxpooling to each cell output to reduce the feature dimension by one-half. This simple, yet efficient solution helps the model to learn a more compact, high-level representation, without increasing the number of channels, thereby reducing computational complexity and demands upon GPU memory.  An added benefit is that the same architecture depth and initial number of channels can be used for both architecture search as well as train from scratch stages. The so-called \emph{depth gap}~\cite{chen2019pdarts, yang2019evaluation} is therefore avoided, where the searched operations may not fit the deeper network in the second stage due to the depth mismatch between architecture search and train from scratch stages.  Thus, the cells used in Raw PC-DARTS are referred to as a \emph{normal} cell and an \emph{expand} cell.  Both cells halve the input feature dimension, whereas only the expand cell doubles the number of channels. Expand cells are placed at the same network depth as reduction cells in the original DARTS approach.
    
    \subsection{Embedding extraction and loss function}
    
    Frame-level representations produced by the final cell are fed to a gated recurrent unit (GRU) layer to obtain  utterance-level representations.  These representations are then fed to a fully connected layer which extracts the embedding. We use mean-square error (MSE) for P2SGrad~\cite{wang2021comparative} as the loss function. An activation layer is first applied to calculate the cosine distance $\cos\theta$ between the input embedding and the class weight. As in~\cite{zhang2019p2sgrad}, this step is hyper-parameter-free, which reduces the sensitivity of margin-based softmax towards its scale and angular margin parameter settings, thus giving relatively consistent results. The network loss is the MSE between $\cos\theta$ and the target class label. Scores used for performance evaluation are $\cos\theta$ for the bona fide class.

\section{Experiments}
    
    \subsection{Database and metrics}
    All experiments were performed using the ASVspoof 2019 Logical Access (LA) database~\cite{wang2020asvspoof} which comprises three independent partitions: train, development and evaluation.  
        Each partition is used in the same way reported in~\cite{ge2021pc-darts}. During architecture search,  network parameters are updated using 50\% of the bona fide utterances and 50\% of the spoofed utterances in the training partition.  Remaining data is used to update architecture parameters. The cell architectures are selected from those which give the best classification accuracy for the full development partition. During the train from scratch stage, all network parameters, except those of the first convolutional layer, are updated using the full training partition and the best model is selected according to that which gives the best classification accuracy for the full development partition. We report results according to two different metrics: the pooled minimum normalised tandem detection cost function~(min-tDCF)~\cite{kinnunen2020tdcf}; the pooled equal error rate~(EER).
        
    \subsection{Implementation details}
    
    We experimented with 3 different sinc filter frequency scales: Mel, inverse-Mel and linear~\cite{tak2021rawnet2}. We tested two settings in each case, namely \emph{fixed} and \emph{learnable}. Fixed scales are set and left unchanged for both architecture search and train from scratch stages. Learnable scales are initialised in the same way, but the configuration is updated during architecture search.  They are then fixed and left unchanged during the train from scratch stage. We also tested a randomly initialised, learnable convolution block denoted Conv\_0, in place of sinc filters. The kernel size, stride and the number of output channels for the Conv\_0 system are set to the same as that of systems that use sinc filters. The maximum number of masked filters is set to $F=16$.
    
    Following~\cite{ge2021pc-darts}, the number of nodes in each cell is fixed to $N=7$ and the number of intermediate node inputs is fixed to 2. Models comprise 8 cells (6 normal cells and 2 expand cells) with $C=64$ initial channels in both stages. During architecture search, we perform 30 epochs of training. In the first 10 designated warm-up epochs, only network parameters are updated.  Both architecture parameters and network parameters are updated in the subsequent 20 epochs. In all cases, the batch size is set to 14 and learning is performed using Adam optimisation. Architecture parameters are updated using a learning rate of 6e-4 and a weight decay of 0.001. Network parameters are updated using a learning rate of 5e-5. Partial channel selection is performed with a value of $K_C=2$. During the train from scratch stage, all models are trained for 100 epochs with a batch size of 32. The initial learning rate of 5e-5 is annealed down to 2e-5 following a cosine schedule. 
    
    All models reported in this paper are trained once with the same random seed on a single NVIDIA GeForce RTX 3090 GPU.  Architecture search takes approximately 21.5 hours, whereas the train from scratch process takes approximately 9.5 hours.  Results are reproducible with the same random seed and GPU environment using the implementation available online\footnote{\href{https://github.com/eurecom-asp/raw-pc-darts-anti-spoofing}{https://github.com/eurecom-asp/raw-pc-darts-anti-spoofing}}.
    
\section{Results}

    First we report a set of experiments which assess the performance of Raw PC-DARTS when using different first layer sinc filter scales.  Next, we present a comparison of performance to existing state-of-the-art solutions.  Finally, we present an analysis of generalisability in terms of performance stability across different spoofing attacks.
    
    \subsection{Raw PC-DARTS with different sinc scales}
    
    Table~\ref{tab:sinc scales_2} shows results in terms of both the min t-DCF and EER for the ASVspoof 2019 LA evaluation partition.  Results are shown for four different sinc scale configurations: Mel; inverse-Mel; linear and with randomly initialised, learnable convolution blocks --- Conv\_0.  With the exception of Conv\_0, results in each case are shown for both fixed and learnable configurations.  
    
    The lowest min t-DCF of 0.0517 (EER of 1.77\%) is obtained using fixed Mel scale sinc filters.  For both inverse-Mel and linear scales, learnable configurations give better results than fixed configurations, with the second best result with a min t-DCF of 0.0583 (2.1\%) being achieved using a linear scale. While the Conv\_0 system achieves a respectable EER of 2.49\%, the min t-DCF of 0.0733 is notably worse than that of the better performing configurations.
    
\begin{figure}[!h]
  \begin{subfigure}{\columnwidth}
    \includegraphics[scale=0.2]{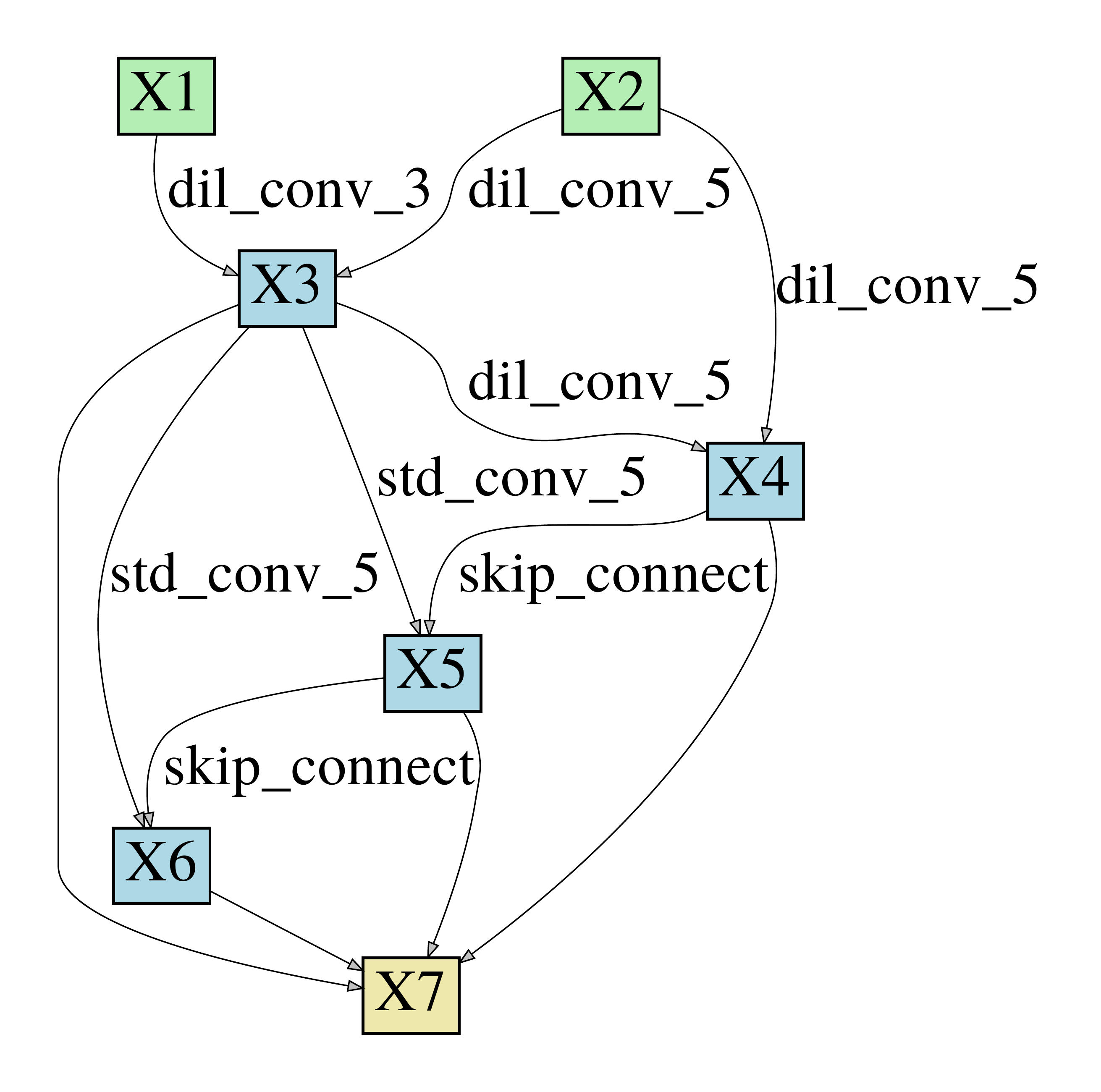}
    \centering
\vspace{-0.25cm}
    \caption{Normal cell}
    \label{fig:normal_cell_1}
  \end{subfigure}
  \hfill 
  \begin{subfigure}{\columnwidth}
  \includegraphics[width=\columnwidth, scale=1.3]{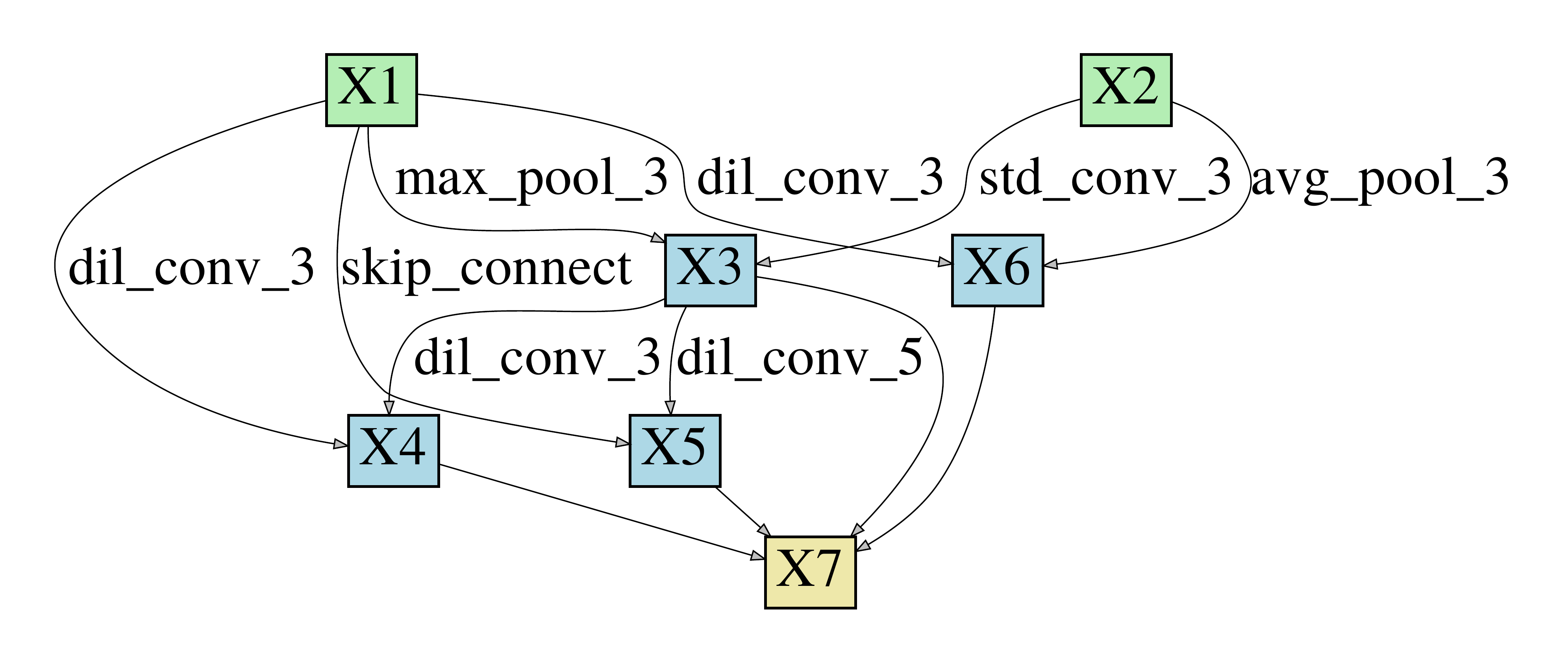}
  \vspace{-0.25cm}
    \caption{Expand cell}
    \label{fig:expand_cell_1}
  \end{subfigure}
  \centering
  \caption{An illustration of the normal (a) and expand (b) cells produced by the architecture search stage for the Mel-Fixed Raw PC-DARTS configuration.}
  \label{fig:cellIllustrations_1}
\end{figure}

    The cell architectures for the best configuration (Mel-Fixed) is illustrated in Fig.~\ref{fig:cellIllustrations_1}.  We observed that, even though architecture parameters are randomly initialised, after several warm-up epochs, those for dilated convolution operations tend to dominate. This may indicated that, compared to other candidate operations within the search space, dilated convolutions contribute more to representation learning when applied to raw waveforms. Dilated convolutions act to increase the receptive field~\cite{hua2021TSSDNet, Tan2018GRUspeech, yu15Dilated}.  The use of greater contextual information then helps to improve performance.
    
    \begin{table}[!t]
         \caption{EER results for the ASVspoof 2019 LA database, evaluation partition.  Results shown for different Raw PC-DARTS setups using different first layer sinc scale initialisations.}
         \label{tab:sinc scales_2}
         \centering
         \renewcommand{\arraystretch}{1.1}
         \setlength\tabcolsep{2pt}
        \begin{tabular}{lccccc}
        \toprule
        \multicolumn{1}{l}{}   &  \multicolumn{2}{c}{Fixed}  & \multicolumn{2}{c}{Learnable}\\
        \multicolumn{1}{l}{\textbf{Type}}  &  \multicolumn{1}{c}{\textbf{min-tDCF}} & \multicolumn{1}{c}{\textbf{EER}} & \multicolumn{1}{c}{\textbf{min-tDCF}} & \multicolumn{1}{c}{\textbf{EER}} \\
        \hline
        \multicolumn{1}{l}{Mel}& \multicolumn{1}{c}{0.0517}   &   
        \multicolumn{1}{c}{1.77}  &  \multicolumn{1}{c}{0.0899}  &  \multicolumn{1}{c}{3.62}   \\
        \multicolumn{1}{l}{Inverse-Mel}& \multicolumn{1}{c}{0.0700}   & 
        \multicolumn{1}{c}{3.25}  &  \multicolumn{1}{c}{0.0655}  &  
        \multicolumn{1}{c}{2.80}        \\
        \multicolumn{1}{l}{Linear} &  \multicolumn{1}{c}{0.0926}  &  
        \multicolumn{1}{c}{3.29} & \multicolumn{1}{c}{0.0583}              & 
        \multicolumn{1}{c}{2.10}         \\
        \multicolumn{1}{l}{Conv\_0} &  \multicolumn{1}{c}{$\times$} & \multicolumn{1}{c}{$\times$}  & \multicolumn{1}{c}{0.0733}  &    
        \multicolumn{1}{c}{2.49}  \\
        
        \bottomrule
        
        \end{tabular}
        \end{table}
        
    \subsection{Comparison to competing systems}
    \begin{table*}[!t]
        \caption{A performance comparison between proposed models and competing state-of-the-art systems reported in the literature.  Results for the ASVspoof LA evaluation partition. }
      \label{tab:baselines}
      \centering
      \renewcommand{\arraystretch}{1.1}
      \setlength\tabcolsep{3pt}
      \begin{tabular}{l c c c c c c c}
        \toprule
        {\textbf{Systems}}&{\textbf{Features}} &{\textbf{min-tDCF}}&{\textbf{EER}}&{\textbf{Params}}&{\textbf{Worst attack}}&{\textbf{Worst EER}}\\
        \midrule
         Res-TSSDNet~\cite{hua2021TSSDNet} & waveform & 0.0482 & 1.64 & 0.35M  & A17 & 6.01\\

        \textbf{Raw PC-DARTS Mel-F} & waveform & 0.0517 & 1.77 & 24.48M  & A08 & 4.96 \\
         
        ResNet18-LCML-FM~\cite{chen2020generalization} & LFB & 0.0520 & 1.81 & -  & A17 &  6.19 \\
         
        LCNN-LSTM-sum~\cite{wang2021comparative} & LFCC & 0.0524 & 1.92 & 0.28M & A17  & 9.24  \\
         
        Capsule Network~\cite{Luo2021Capsule} & LFCC & 0.0538 & 1.97 & 0.30M & A17 & 3.76 \\
         
        \textbf{Raw PC-DARTS Linear-L} & waveform & 0.0583 & 2.10 & 24.40M & A08 & 6.23  \\
        
        ResNet18-OC-Softmax~\cite{zhang2021oneclass} & LFCC & 0.0590 & 2.19 & -  & A17 & 9.22 \\
         
        Res2Net~\cite{li2020res2net}&CQT &0.0743 & 2.50& 0.96M & - & -\\
         
        ResNet18-AM-Softmax~\cite{zhang2021oneclass} & LFCC & 0.0820 & 3.26 & -  & A17 & 13.45\\
        
        ResNet18-GAT-T~\cite{tak2021GAT} & LFB & 0.0894 & 4.71 & -  & A17 & 28.02\\
        
        ResNet18-GAT-S~\cite{tak2021GAT} & LFB & 0.0914 & 4.48 & -  & A17 & 21.74\\
        
        PC-DARTS~\cite{ge2021pc-darts}&LFCC &0.0914 & 4.96& 7.51M  & A17 & 30.20\\
        
        RawNet2~\cite{tak2021rawnet2} & waveform & 0.1294 & 4.66 & 25.43M  & A18 & 16.30 \\
        \bottomrule
      \end{tabular}
    \end{table*}
    
    Table~\ref{tab:baselines} shows a comparison of results for the two best performing Raw PC-DARTS systems to that of the top-performing systems reported in the literature\footnote{Number of learnable parameters and the decomposed EER results for Res-TSSDNet and LCNN-LSTM-sum were obtained using open-source codes available online. Those for Capsule Network were provided by the authors of~\cite{Luo2021Capsule}, those for ResNet18-GAT and RawNet2 were provided by the authors of~\cite{tak2021rawnet2, tak2021GAT}.}.  Among the illustrated systems, four operate upon raw inputs, including the top two systems, the first of which is the Res-TSSDNet system reported in~\cite{hua2021TSSDNet} and the second of which is the proposed Raw PC-DARTS. The fourth system which operates on the raw waveform is the RawNet2 system reported in~\cite{tak2021rawnet2}.  It also uses a first layer of sinc filters, GRU and fully connected layer for embedding extraction.  
    
    These results point toward the competitiveness of solutions that operate upon the raw waveform but also show that solutions whose cell architectures are learned automatically can perform almost as well or better that those that are hand-crafted.
    
    \subsection{Complexity}
    
    The number of network parameters for the systems illustrated in Table~\ref{tab:baselines} is shown in column 5 (where such numbers are available). The two best Raw PC-DARTS architectures have in excess of 24M parameters. For the Mel-Fixed configuration, 77\% (18.89M) of the learnable network parameters correspond to GRU layers wereas only 18\% (4.52M) correspond to the stacked cells. The RawNet2 system, which also uses a GRU, has over 25M parameters. Other systems have far fewer parameters, including the top Res-TSSDNet system which has 0.35M parameters.  It uses ResNet-style 1D convolution blocks and 3 FC layers, without GRUs. The use of dilated convolutions helps to control network complexity while increasing the receptive field~\cite{hua2021TSSDNet}. Though the LCNN-LSTM-sum system uses two bidirectional LSTM layers, which is normally computationally expensive, use of a hidden size of 48 nonetheless means that the complexity is the lowest of all illustrated systems.  The additional complexity of the Raw PC-DARTS architecture is currently a limitation in the approach, yet a compromise that might be acceptable given that learning and optimisation is a one-step process requiring comparatively little human effort.
    
    \subsection{Worst case scenario}
    
    Generalisation has been focus of anti-spoofing research since the inception of the ASVspoof initiative.  It is well known that even top-performing systems can struggle to detect the full range of spoofing attacks~\cite{nautsch2021asvspoof}.  There is hence interest in minimising not just pooled performance, but also that for the so-called \emph{worst case scenario} which, for the ASVspoof 2019 LA database, is generally the infamous A17 attack.  
    
    The worst case attack and corresponding EER for each system is shown in columns 6 and 7 of Table~\ref{tab:baselines}.  Here we see a distinct advantage of systems that operate upon raw inputs.  The Res-TSSDNet~\cite{hua2021TSSDNet} and both Raw PC-DARTS systems have among the lowest worse case EERs. This observation indicates that the waveform based systems can capture discriminative artefacts that are missed by systems that use hand-crafted inputs. Were an adversary to discover the attacks to which a system is most vulnerable and exploit only attacks of this nature, then the Raw PC-DARTS countermeasures would offer the second-best protection among all competing systems. 
    
\section{Conclusion}

    In this paper, we proposed an end-to-end differentiable architecture search approach to speech deepfake and spoofing detection, named Raw PC-DARTS. We show that the components of a deep network model, including pre-processing operations, network architecture and parameters, can all be learned automatically from raw waveform inputs and that the resulting system is competitive with the state of the art.
    
    While the best performance is obtained using a fixed front-end, rather than with a learnable configuration, the latter is only marginally behind, while both systems give among the best performance reported to date for the ASVspoof 2019 logical access database.  The use of gated recurrent units means that the resulting models are, however, substantially more complex than competing systems and may exhibit some redundancies. While it may be possible to reduce redundancy, and while the results reported in the paper are the first to show the genuine potential of learned architectures, further work to tackle complexity is required if they are to be competitive when computational capacity is limited and a design criteria, e.g.\ for embedded applications. 
    One avenue for future research in this direction is to evaluate the replacement of gated recurrent units, with a number of parameters in the millions, with concatenated fully connected layers with orders of magnitude fewer parameters.
    
    We also observe that the Raw PC-DARTS solution generalises better to unseen forms of spoofing attacks than their hand-crafted counterparts.  Performance for the worst case A17 attack is notably better than that for competing systems.  We are currently working to understand what information or cues missed by handcrafted solutions are captured successfully by fully learned solutions.  With answers to these questions, we may be able to combine the benefits of both in order to improve reliability further while also protecting complexity.
    
\section{Acknowledgements}

    This work is supported by the TReSPAsS-ETN project funded from the European Union’s Horizon 2020 research and innovation programme under the Marie Skłodowska-Curie grant agreement No.860813. It is also supported by the ExTENSoR project funded by the French Agence Nationale de la Recherche (ANR).

\bibliographystyle{IEEEbib}
\bibliography{ASVspoof2021_BibEntries}
\end{document}